# Modular 3D Interface Design for Accessible VR Applications

Corrie Green[1]

Robert Gordon University

c.green1@rgu.ac.uk

Dr John, Isaacs

Dean, Robert Gordon University

j.p.isaacs@rgu.ac.uk

January 26th, 2023

## ABSTRACT

Designed with an accessible first design approach, the presented paper describes how exploiting humans' proprioception ability in 3D space can result in a more natural interaction experience when using a 3D graphical user interface in a virtual environment. The modularity of the designed interface empowers the user to decide where they want to place interface elements in 3D space allowing for a highly customizable experience, both in the context of the player and the virtual space. Drawing inspiration from today's tangible interfaces used, such as those in aircraft cockpits, a modular interface is presented taking advantage of our natural understanding of interacting with 3D objects and exploiting capabilities that otherwise have not been used in 2D interaction. Additionally, the designed interface supports multimodal input mechanisms which also demonstrates the opportunity for the design to cross over to augmented reality applications. A focus group study was completed to better understand the usability and constraints of the designed 3D GUI.

## 1  Extended Abstract

The progression of the GUI has reached a level of ubiquitous understanding so that many users can use a computer or mobile device and understand how to interact with windows, icons, menus, and pointers on a 2D interface.

However, with the adoption of virtual reality, these design decisions have for the most part been directly translated to work in a 360 6-dof experience. Resulting in the same 2D windows icons menus and pointers being used in VR as 2D displays. With the VR interaction mechanism of choice being a ray cast in the form of a laser, designers can interact with UI elements the same way a cursor would on a desktop environment. Thus, reducing the development time to implement a new interface.

By relying on traditional interface design decisions, we are not necessarily taking advantage of full 360 6-dof environments that we see in the real world. An extra dimension of interaction is made available, but in many VR scenarios designers are defaulting to using or imagining a 2D WIMP interaction technique for a virtual reality play space. Existing interaction techniques for virtual reality have resulted in novel and diverse set of designs that has primarily been explored in the VR games market[1].

---

[1] https://orcid.org/0000-0003-0404-3668.

3D interfaces can utilize our sense of proprioception as we physically surround ourselves with static interfaces in the form of 3D objects. Proprioception can also be referred to as kinaesthesia and is the ability to know where your body is in space and has been referred to as our "six sense" [2].  It gives the user an idea of where their body parts are in relation to their environment and how they may be able to interact with it. Drawing closer to incorporating the human body fully can increase our understanding of the 3D digital environment presented, by using our natural understanding of the world [3].

As VR has the potential to provide more presence than a traditional interface, the idea requires more interface design evolution to become more flexible and human oriented. An example study simulated wind, the breeze enhanced presence among users [4]. Showing that somewhat unconventional methods can enhance presence, to which, a modular interactive interface for the user may furthermore provide fewer constraints leading to a more natural interaction technique. By drawing inspiration from the tangible interfaces, we use today - such as those in aircraft cockpits - a modular 3D GUI has developed taking advantage of human proprioception capabilities which otherwise have not been fully exploited for interacting with interfaces. The developed system is compatible with existing ray-based interaction approaches, reducing the adoption effort to users who may have accessibility constraints. The system allows for many standard practices from 2D wimp interface design to be migrated across but in a new medium of 3D panels over windows or forms. Instead, gesture-based interaction approaches have been used which allow the users' hands to be tracked with or without a controller to use for selection and navigation. The system is modular allowing users to move the location of all UI components in 3D space to suit their preference, with the addition of contextual-based UI being explored where UI elements appear only when relevant to the task being completed.

An important aspect to consider before exploring the opportunities made by a 3D GUI was the vehicle in which to showcase the interface itself. Usability discussions were completed in focus groups which allowed participants to share their experience using the developed interface alongside peers and provide insight into the satisfying and frustrating aspects of the interaction. Establishing a familiar metaphor was important as recruitment was focused on users' previous exposure to virtual reality applications, where novice to intermediate users were selected to participate. The selected demographic was used as the study's focus was to understand the usability of the designed interface and highlight issues and struggles with a 3D modular GUI.

A conceptual model was therefore established to understand the problem space for a new medium of interaction. To present the interface in an application that a 3D GUI may be applicable to, the development of a novel 3D visualization platform for geospatial positions was developed. Allowing for visualization of various forms of 3D spatial data including point cloud scans, video playback and geospatial data which was represented by electric vehicle (EV) chargers' latitude and longitude positions. Querying of EV charger locations was completed based on their charging type and availability. This allowed for a selection of chargers to be scanned and linked to their real-world position visualized by their point cloud scan using the Microsoft Azure Kinect. The map was selected as the primary visualization tool as a homage to the map being the "graphic representation of the milieu" [5] but primarily due to it being a familiar visualization medium for many people.

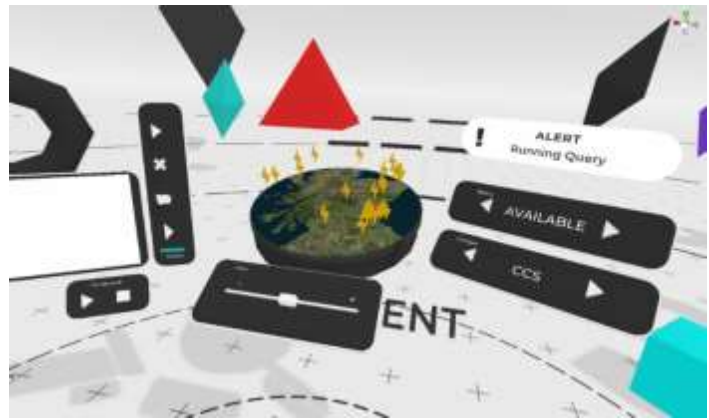
Figure 1: Screenshot of the developed interface allowing for geospatial locations to be represented.

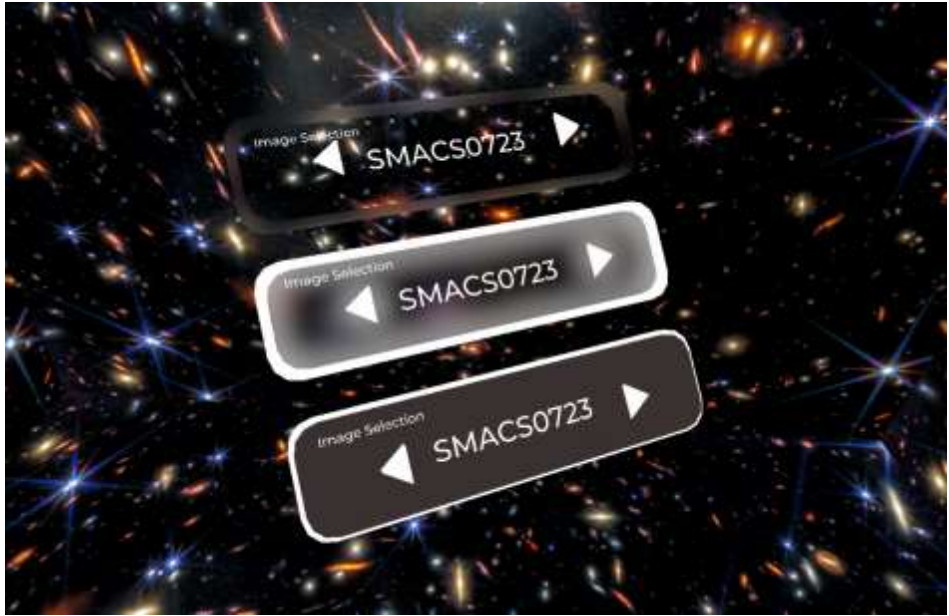

Figure 2: Demonstration of button input on the panels. White elements are grabbable. The buttons support physical spring-based interaction so will respond with motion. Screenshot demonstrates opacity with complex background objects.

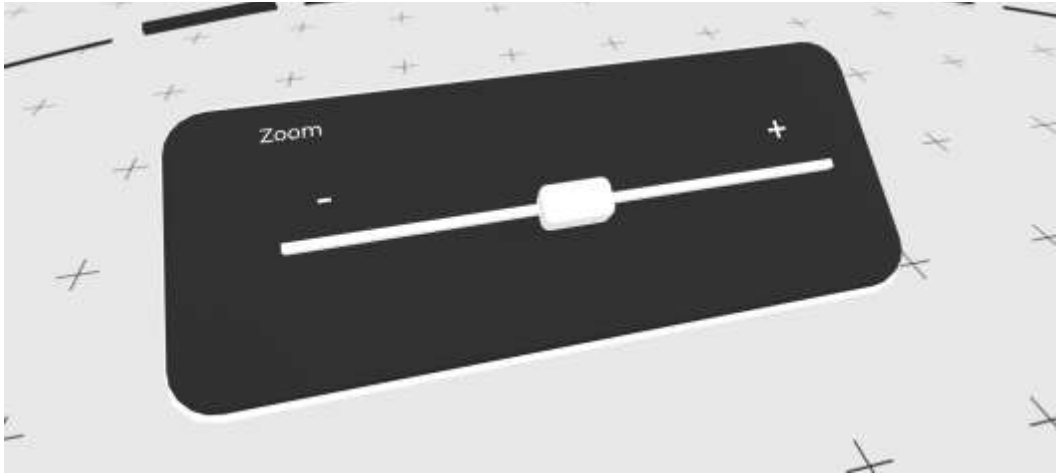

Figure 3: 3D slider which supports a physic spring joint movement, the anchor point is always center, and the user slides left or right to increase scale or decrease as a continuous action.

Latest version of the demo and repo can be found [here](here)